\documentclass[10pt, conference]{IEEEtran}
\usepackage{algorithmicx}
\usepackage[ruled,vlined,linesnumbered]{algorithm2e}
\usepackage{hhline}
\usepackage{amsmath,mathtools}
\usepackage{amsfonts,amssymb}
\usepackage{mathrsfs}
\usepackage{caption}
\usepackage{enumitem}
\usepackage{multirow}
\usepackage{graphicx} 
\usepackage{enumitem,color}
\usepackage{algpseudocode}

\begin{document}

\IEEEoverridecommandlockouts\IEEEpubid{\makebox[\columnwidth]{ 978-1-6654-4131-5/21/\$31.00 $\copyright$2021 IEEE \hfill}\hspace{\columnsep}\makebox[\columnwidth]{ }}
%

\title{\huge Constraint-Aware Deep Reinforcement Learning for End-to-End Resource Orchestration in Mobile Networks \vspace{-0.1in}}

\author{\IEEEauthorblockN{Qiang Liu \vspace{-0.15in}}\\
\IEEEauthorblockA{University of Nebraska-Lincoln\\
qiang.liu@unl.edu}\vspace{-0.3in}
\and
\IEEEauthorblockN{Nakjung Choi \vspace{-0.15in}}\\
\IEEEauthorblockA{Nokia Bell Labs\\
nakjung.choi@nokia-bell-labs.com}\vspace{-0.3in}
\and
\IEEEauthorblockN{Tao Han \vspace{-0.15in}}\\
\IEEEauthorblockA{New Jersey Institute of Technology\\
tao.han@njit.edu}\vspace{-0.3in}
}


\maketitle

\begingroup
\renewcommand\thefootnote{\textsection}
\footnotetext{This work is completed at Nokia Bell Labs. Dr. Tao Han's work is partially supported by the US National Science Foundation under Grant No. 1910844, No. 2008447, No. 2047655, and No. 2049875.}
\footnotetext{We would like to thank the shepherd, Prof. Kate Ching-Ju Lin for the guidance in refining this paper and anonymous reviewers for their insightful comments.}
\endgroup

\begin{abstract}
Network slicing is a promising technology that allows mobile network operators to efficiently serve various emerging use cases in 5G. It is challenging to optimize the utilization of network infrastructures while guaranteeing the performance of network slices according to service level agreements (SLAs). To solve this problem, we propose \emph{SafeSlicing} that introduces a new constraint-aware deep reinforcement learning (CaDRL) algorithm to learn the optimal resource orchestration policy within two steps, i.e., offline training in a simulated environment and online learning with the real network system. On optimizing the resource orchestration, we incorporate the constraints on the statistical performance of slices in the reward function using Lagrangian multipliers, and solve the Lagrangian relaxed problem via a policy network. To satisfy the constraints on the system capacity, we design a constraint network to map the latent actions generated from the policy network to the orchestration actions such that the total resources allocated to network slices do not exceed the system capacity. We prototype \emph{SafeSlicing} on an end-to-end testbed developed by using OpenAirInterface LTE, OpenDayLight-based SDN, and CUDA GPU computing platform. The experimental results show that \emph{SafeSlicing} reduces more than $20\%$ resource usage while meeting SLAs of network slices as compared with other solutions. 
\end{abstract}


\section{Introduction}
\label{sec:introduction}
5G is designed to support three primary use cases, i.e., enhanced Mobile Broadband (eMBB), Ultra Reliable Low Latency Communications (URLLC) and Massive Machine Type Communication (mMTC), which consequently enable various new applications such as augmented/virtual reality (AR/VR), 360-degree video streaming, and vehicle-to-everything (V2X)~\cite{wp5d2017minimum}.
These new applications have diverse requirements of quality of services (QoS), e.g., data rates, latency, jitters, and reliability. 
It is thus desirable to cost-efficiently customize mobile networks and provision networking and computing resources for individual applications according to their demands~\cite{afolabi2018network}.  

\begin{figure}[!t]
	\centering	
	\includegraphics[width=3.4in]{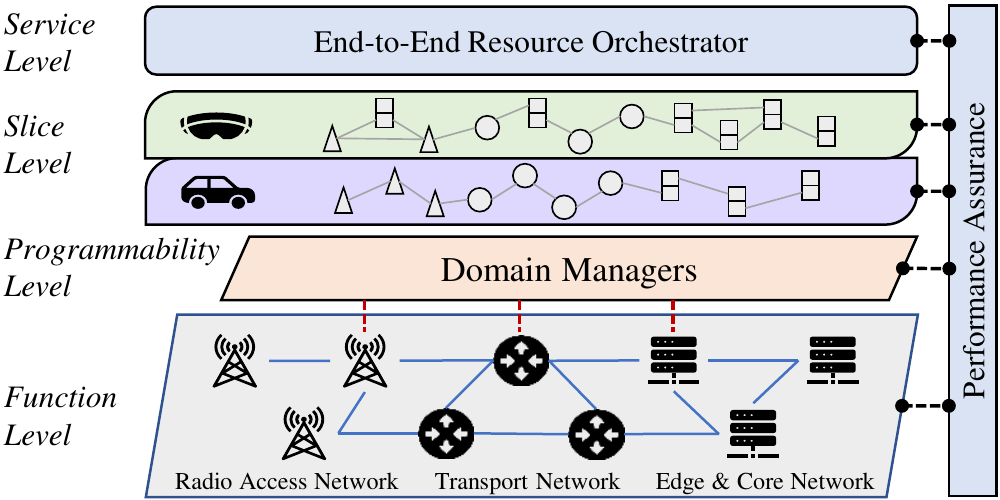}\vspace{-0.05in}
	\caption{\small The architecture of network slicing management}
	\label{fig:network_archi}
\end{figure}

Network slicing allows mobile network operators (MNOs) to virtualize physical network infrastructures, e.g., base stations and servers, and provide network slices with isolated resources to slice tenants~\cite{foukas2017network}.
A network slice is an end-to-end virtual network customized to fulfill the QoS requirements of a particular application based on the service level agreement~(SLA)~\cite{ordonez2017network}.
To accommodate more slices simultaneously and thus improve the network revenue, MNOs aim to satisfy the QoS requirement of slices with the minimum resource usage in multiple technical domains~\cite{salvat2018overbooking}.
As shown in Fig.~\ref{fig:network_archi}, the slice management~\cite{3gpp-nsmf} is composed of multiple levels, e.g., function, programmability, slice and service levels.
At the programmability level, multiple domain managers are developed to enable the abstraction of network functions (NFs), e.g., physical, virtual and containerized NFs, and implement the orchestration decision from the resource orchestrator.
An end-to-end resource orchestrator at the service level dynamically orchestrates the virtual resources (e.g., virtual UL radio resource) to diverse slices for meeting their performance requirements.
It observes high-dimension network states from domain managers and slice tenants, and makes orchestration decisions to assure performance of slices.

It is challenging to orchestrate cross-domain resources due to the complicated interdependencies in end-to-end network systems.
On one hand, the slice performance is correlated to many distinct types of resources, e.g., bandwidth in transport network and CPU/IO in edge servers, where an accurate mathematical model can hardly be derived~\cite{liu2020edgeslice, zhang2020onrl}.
Consequently, model-based approaches fail to precisely represent the complex network slicing system and might result in performance degradation if applied.
On the other hand, the varying high-dimensional network states also affect the slice performance.
For example, given a resource orchestration, the round-trip application latency of an AR/VR slice may change under different radio channel qualities and traffic of slice users.
In addition, the resource orchestration problem shows the Markov property, i.e., the next slice performance and network states depend only upon the current orchestration decision and observed network states such as service queuing in edge servers.
Therefore, a model-free solution that can handle the high-dim interdependencies is needed to orchestrate cross-domain resources in end-to-end network slicing.

Deep reinforcement learning (DRL) has gained increasing popularity in managing and optimizing dynamic and complex network systems in a model-free approach~\cite{liu2020edgeslice,mao2019learning,ayala2019vrain}, and achieves promising performance improvements by parameterizing policies with the deep neural networks (DNNs).
However, most of the existing DRL solutions derive their policies within simulated environments and directly apply them into real networks without any adaptations.
These offline simulated environments are usually built by leveraging approximated models, e.g., queuing, which cannot completely represent the complex network systems, especially in end-to-end domains.
In other words, there is a discrepancy between the offline environment and the real network~\cite{mao2019learning, zhang2020onrl}, which is not negligible in practice and cannot be overlooked.
As a result, deploying the offline-trained policy for resource orchestration in real networks could cause performance degradation of slices and thus lead to the violation of slice SLAs.
Thus, an online DRL solution is indispensable for eliminating the simulation-to-reality discrepancy, which allows the policy to be learned by directly interacting with a real network.


On slicing end-to-end networks, DRL needs to be constraint-aware for two main reasons.
First, the policy should satisfy the statistical performance requirement derived from slice SLAs, e.g., 95-percent end-to-end application latency in a day.
Reward-shaping methods are proposed to solve this problem by shaping the reward function with static weighted constraints. However, these methods may result in either violations of SLAs when the weights are too small or the suppression of exploitation when the weights are too large.
Second, the policy needs to maintain the instantaneous system limitations at any time slots, e.g., the summation of the bandwidth allocated to all slices cannot exceed its total link bandwidth.
On learning the orchestration policy, a DRL algorithm needs to explore the whole action space containing all possible actions, i.e., resource orchestration of slices, for seeking better rewards.
For example, a widely-adopted random action exploration strategy adds small deviations, which are sampled from a Normal distribution, onto the actions generated by the policy.
As these deviated actions are implemented in the real network, multiple system limitations could be breached.

In this paper, we propose \emph{SafeSlicing} that allows MNOs dynamically slice end-to-end network system, i.e., optimizes the cross-domain resource usage while guaranteeing the performance requirements of slices and maintaining the system limitations.
\emph{SafeSlicing} is accomplished by a novel constraint-aware deep reinforcement learning (CaDRL) algorithm in two steps, i.e., offline training in a simulated environment and online learning with the real network.
First, we develop a simulated environment to imitate the resource orchestration in the real network, by using real experimental dataset and exploiting domain knowledge such as queuing service model.
The environment is designed to reduce the simulation-to-reality gap, and used to offline train the policy, which thus boosts the policy performance before online learning.
Second, we allow the policy to be refined in an online learning manner by interacting with the real network directly.
The policy can quickly adapt to the real network with the CaDRL algorithm and hence eliminate the simulation-to-reality discrepancy.


%
The CaDRL algorithm incorporates the statistical performance constraints in its reward function by using Lagrangian primal-dual method and uses a novel constraint neural network as the output layer appending to the end of the policy network to regulate the resource orchestration actions.
This constraint network ensures all the actions generated by the policy satisfy the constraints on instantaneous resource allocations at any time slots.
Because of the constraint awareness, CaDRL can directly interact with the real network and learn the resource orchestration policy.
We validate the performance of the CaDRL algorithm and prototype \emph{SafeSlicing} on an end-to-end network slicing testbed designed with OpenAirInterface, OpenDayLight SDN, and CUDA GPU computing platform\footnote{\emph{SafeSlicing} fits well with 5G architecture that provides open interfaces and flexible control functions such as O-RAN (open RAN)~\cite{oran}, NWDAF (NetWork Data Analytics Function), and NEF (Network Exposure Function).}.

The contributions of this paper are summarized as follows:
\begin{itemize}
    \item We design \emph{SafeSlicing}, the first integrated offline-online learning system, that can dynamically orchestrate networking and computing resources for end-to-end network slicing and optimize the resource utilization of network infrastructures while meeting the SLAs of slices. 
    \item We develop a novel constraint-aware deep reinforcement learning (CaDRL) algorithm that can learn the orchestration policy directly from the real network. CaDRL can effectively handle the constraints on the statistical slice performance and instantaneous resource allocation during the policy optimization so that the violations of slices' SLAs approximate to zero.
    \item We prototype \emph{SafeSlicing} on a small-scale end-to-end network slicing testbed, by using OpenAirInterface LTE to implement the radio access network, SDN with the OpenDayLight controller to emulate the transport network, and CUDA GPU computing platform for computing acceleration.
    \item We conduct extensive experiments to evaluate the performance of \emph{SafeSlicing}. The experimental results show that \emph{SafeSlicing} outperforms other existing solutions in terms of resource usage and slice performance.
\end{itemize}

\begin{figure*}[!t]
	\centering
	\includegraphics[width=5.8in]{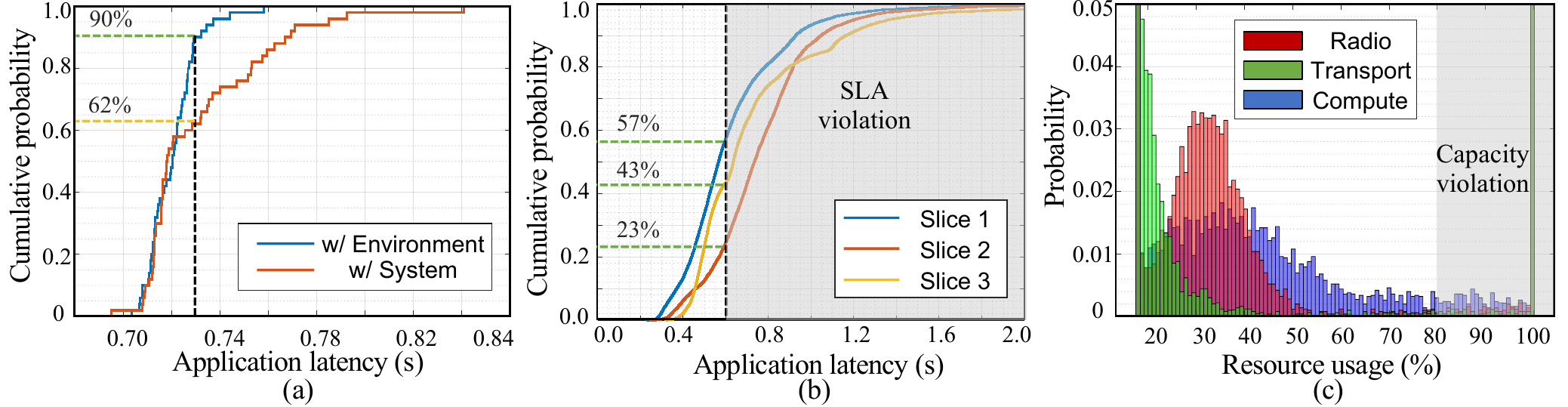}\vspace{-0.05in}
	\caption{\small a) CDF of app. latency within system and simulated environment under identical policy; b) CDF of app. latency of slices during online learning with reward shaping; c) the probability distribution of resource usages during online learning with reward shaping.}
	\label{fig:motivations}
\end{figure*}
\section{Background and Motivation}

With dynamic network slicing, the MNO aims to minimize the usage of cross-domain resources while satisfying the statistical performance requirements of network slices with the capacity constraints on physical networking and computing infrastructures. 
For example, an AR/VR slice needs uplink radio resources to upload environment sensing data, transport resources to transmit data to edge servers, computing resources to process data, and downlink radio resources for downloading augmentation contents.
To accomplish this slice's SLA, e.g., 500 ms average round-trip latency, multiple resource orchestrations may be feasible, e.g., 10 UL/DL PRBs in RAN, 100Mbps bandwidth in transport network, and 1 dedicated GPU in the edge server.
However, the minimal resource usage closely depends on the high-dimensional network state, e.g., slice traffic, radio channel quality, transport congestion, and service queuing in servers.
For instance, this slice requires more UL/DL PRBs to maintain a similar performance if the radio channel quality of slice users is poor, as each PRB carries less data.
Besides, the capacity of resources are constrained by physical network infrastructures, e.g., the total number of PRBs, and thus the orchestration of all slices need to be jointly optimized in case of resource over-requesting.
Therefore, we resort to the deep reinforcement learning (DRL) technique to deal with complicated interdependencies in the complex end-to-end slicing system.

Applying DRL in dynamic network slicing faces two challenges.
The first one is the discrepancy between the simulated environment and real networks, which causes that a DRL policy trained offline within a simulated environment may not work well in real networks. The second one is the constraint violation during the online learning as the intrinsic DRL exploration mechanism.


\textbf{Simulation-to-reality discrepancy}.
To study the impact of the discrepancy, we develop an end-to-end network slicing testbed (detailed in Sec.~\ref{sec:implementation}) and build a simulated environment that mimics the resource orchestration in the real network.
{We train the DRL policy in the simulated environment and then apply the policy in the real network without online learning.}
Fig.~\ref{fig:motivations} (a) shows the performance comparison of a resource orchestration policy in the simulated environment and system testbed.
In the simulated environment, the policy obtains the application latency of slices range  from $0.7s$ to $0.76s$ with about $90\%$ of it being less than $0.73s$.
In contrast, when the same policy is applied in the system testbed, the application latency of slices range from $0.69s$ to $0.83s$, and only $62\%$ of it is less than $0.73s$. 
This result shows the performance degradation when applying an offline-trained policy to a real network.
This simulation-to-reality discrepancy also exists in various systems, e.g., data center networks \cite{mao2019learning}, robot system~\cite{tan2018sim}, and video telephony~\cite{zhang2020onrl}, and can be hardly eliminated as real networks are too complicated to be precisely represented by simulated environments.


\textbf{Constraint violation}. The problem caused by the simulation-to-reality discrepancy requires us to design an online DRL approach that can directly interact with and learn the network slicing policy from real networks.
However, during the online learning, the DRL agent may violate the SLAs and system capacity constraints. 
Assume that the maximum allowable application latency of a slice is $0.6s$, and the resource usage of all slices cannot exceed $80\%$ of the capacity of individual resources, where the remaining $20\%$ resources are reserved for conventional non-slicing users. 
Fig.~\ref{fig:motivations} (b) and (c) show the performance of an online DRL algorithm that uses the reward shaping technique to shape the reward function with fixed weighted constraints~\cite{laud2004theory}.
In Fig.~\ref{fig:motivations} (b), it shows that the probability of satisfying the maximum latency constraint in slice 1, 2, and 3 are only 57\%, 23\% and 43\%, respectively.
In Fig.~\ref{fig:motivations} (c), we can see that there are many instances in which the resource usages exceed $80\%$ of the capacity of individual resources.

These experiment results demonstrate the need of a safe network slicing system that enables MNOs to dynamically optimize the resource orchestration among slices without violating the SLA and capacity constraints.

\section{\emph{SafeSlicing} Overview}
\label{sec:SafeSlicing}

\begin{figure}[!t]
	\centering
	\includegraphics[width=3.45in]{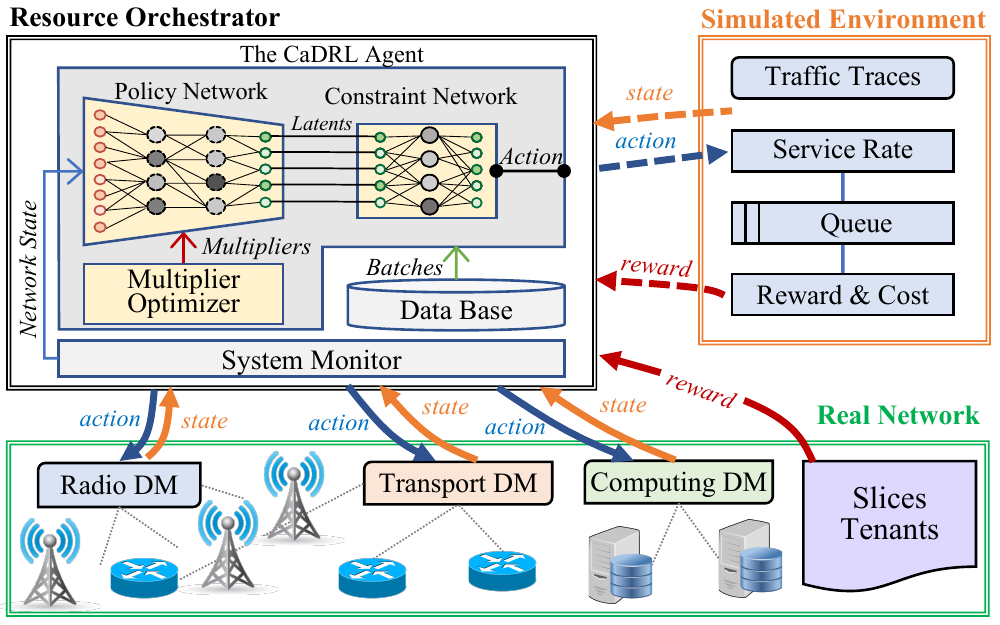}\vspace{-0.05in}
	\caption{\small The \emph{SafeSlicing} system.}
	\label{fig:system_design}
\end{figure}

As shown in Fig.~\ref{fig:system_design}, \emph{SafeSlicing} consists of a resource orchestrator, a simulated environment, and the real network\footnote{\emph{SafeSlicing} is aligned with industry efforts on network slicing, e.g., ETSI ZSM~\cite{zsm}, 3GPP Slice LifeCycle Management (LCM)~\cite{3gpp}.}.
The resource orchestrator centrally manages the allocation of virtual resources in radio access network, transport network, and edge servers, to all slices.
The simulated environment is designed to imitate the resource orchestration in the real network for training the CaDRL agent offline.
The domain managers (DMs) implement the resource orchestration issued by the orchestrator in radio access network, transport network, and edge servers, respectively.
The orchestrator is deployed in the core network and DMs are instantiated in proximity to infrastructures, where they exchange data, e.g., state collection and action notification, through socket communication\footnote{The deployment of \emph{SafeSlicing} can be extended in large-scale operational networks to support distributed DMs. The communication between the orchestrator and DMs can be enhanced by adopting standardized architectures (O-RAN~\cite{oran}) and interfaces (ETSI Network Resource Model~\cite{nrm}).}. 
The network state, e.g., slice traffic and queue, is collected by a system monitor from domain managers and slice tenants.
The system monitor also collects the performance of slices (i.e., end-to-end latency) and stores state-action-performance pairs into a database.
The \emph{SafeSlicing} is accomplished in two steps, i.e., offline training in the simulated environment and online learning with the real network, where the offline trained policy serves as the start point of online learning.


The core of the resource orchestrator is the constraint-aware deep reinforcement learning (CaDRL) agent.
It learns the orchestration policy either in the simulated environment or with the real network, and ensures the slice SLAs and system limitations to be adhered during the exploration of the policy. 
The CaDRL agent consists of three main components, i.e., a \emph{policy network}, a \emph{constraint network}, and a multiplier optimizer. 
The \emph{policy network} is responsible for generating latent actions based on the network \emph{states} to maximize the long-term reward.
The statistical constraints, e.g., average application latency of slices, are incorporated in the reward function so that they can be enforced during the policy optimization. 
During the learning, \emph{policy network} is updated by using actor-critic method~\cite{konda2000actor} under dynamic multipliers.
The multipliers are updated by an optimizer that runs at a slower timescale than that of the update of \emph{policy network} to maintain the statistical constraints in the network slicing (detailed in Sec.~\ref{sec:reward_const}). 
The latent actions are then fed into the pre-trained \emph{constraint network} which generates the final resource orchestration actions. The \emph{constraint network} prevents the orchestration actions from violating the instantaneous resource constraints, e.g., the system capacity of cross-domain resources (detailed in Sec.~\ref{sec:action_const}). 

%
%


\section{System Model}
\label{sec:system_model}

We consider an end-to-end mobile network that consists of a cellular base station (BS) in radio access network (RAN), an edge computing server, and a transport network connecting the RAN and the server. Denote $\mathcal{I}$ and $\mathcal{K}$ as the set of slices and multi-domain resources, respectively. $i \in \mathcal{I}$ and $k \in \mathcal{K}$ represent the $i$th slice and $k$th resource, respectively. The end-to-end mobile network hosts multiple slices according to their service level agreements (SLAs) that define the performance requirements, e.g., the maximum latency~\cite{gsma_sgt}. To optimize the performance of slices, the network operator adjusts the resource orchestration in each domain periodically.


With the consideration of temporal interdependencies and various constraints in the end-to-end network slicing, the cross-domain resource orchestration problem can be naturally formulated as a constrained Markov decision process (CMDP) denoted as $\langle \mathcal{S}, \mathcal{A}, \mathcal{R}, \mathcal{C}, \mathcal{P}, \mu \rangle$.
Here, $\mathcal{S}, \mathcal{A}$ are the set of states and actions, and $ \mathcal{R}, \mathcal{C}$ are the reward and cost functions, respectively.
$\mathcal{P}: \mathcal{S} \times \mathcal{A}  \times \mathcal{S}  \rightarrow [0,1]$ is the transition function, and $\mu: \mathcal{S} \rightarrow [0,1]$ is the initial state distribution.
A parameterized resource orchestration policy $\pi_\theta: \mathcal{S} \rightarrow Pr(\mathcal{A})$ is a mapping from states to probability distribution over actions, where $\theta$ is the neural network parameters.

The state $\mathbf{s}_t \in \mathcal{S}$ is observed by the policy $\pi_\theta$, which represents the status of the end-to-end network.
Thus, we define \emph{\textbf{state}} as the combination of following components: 1) the average traffic of slices $[f^{t-1}_{i}, \forall i \in \mathcal{I}]$, which provides useful information about the user traffic at the current time slot $t$;
2) the number of users waiting in the queue of each slice $[l^{t-1}_{i}, \forall i \in \mathcal{I}]$, which describes traffic loads inherited from last time slot $t-1$; 
and 3) the slice performances in last time slot that consists of the number of users served by each slice $[u^{t-1}_{i}, \forall i \in \mathcal{I}]$, the resource usages $[g^{t-1}_{k}, \forall k \in \mathcal{K}]$, and the performance of slices $[y^{t-1}_{i}, \forall i \in \mathcal{I}]$. This information shows how the resource orchestration in last time slot performs.

Based on the observed state, the policy $\pi_\theta$ generates an action, where $\mathbf{a}_t \in \mathcal{A}$ corresponds to the cross-domain resource orchestration for slices.
We define \emph{\textbf{action}} $\mathbf{a}_t \triangleq [ {a}_{i,k}^{t}, \forall i \in \mathcal{I}, k \in \mathcal{K} ]$, where ${a}_{i,k}^{t} \in [0, 1]$ is the $k$th resource allocated to the $i$th slice at time slot $t$.
In the system, we consider four types of resources (see Sec.~\ref{subsec:CaDRL}), i.e., the uplink and downlink physical resource blocks (PRBs) in radio access network, the bandwidth in transport network, and the GPU computing resources in edge servers.
By taking action $\mathbf{a}_t$ to the system under the state $\mathbf{s}_t$, a reward can be obtained from the system. 
We define \emph{\textbf{reward}} function $\mathcal{R}: \mathcal{S} \times \mathcal{A} \rightarrow \mathbb{R}$ as $r_t \triangleq -\sum\nolimits_{i \in \mathcal{I}} \sum\nolimits_{k \in \mathcal{K}}  {a}_{i,k}^{t}$.
By maximizing the long-term rewards achieved by the policy $\pi_\theta$, we are able to minimize the utilization of cross-domain resources.

Meanwhile, the resource orchestration has to satisfy two types of constraints, which are mapped to the cost functions $\mathcal{C}: \mathcal{S} \times \mathcal{A} \rightarrow \mathbb{R}$.
On one hand, the network operator needs to maintain the slice performances specified by SLAs.
Denote ${y}_{i}^{t}$ is the performance of the $i$th slice at time slot $t$, we define the first \emph{\textbf{cost}} as $c_i \triangleq {y}_{i}^{t}, \forall i \in \mathcal{I}$.
Then, we can maintain the SLA of slices by letting 
\begin{equation}
    \mathbb{E}_{\pi_\theta} \left[ \frac{1}{|\mathcal{T}|} \sum\nolimits_{t \in \mathcal{T}} \gamma^t c_i(\mathbf{s}_t, \mathbf{a}_t) \right] \ge {Y}_{i}, \forall i \in \mathcal{I}, \label{rl_const1}
\end{equation}
where ${Y}_{i}$ is the performance requirement of the $i$th slice.
For instance, the $c_i(\mathbf{s}_t, \mathbf{a}_t)$ can be defined as 90-percent application latency of slice users in a slice and ${Y}_{i}$ is 0.5s.

On the other hand, the resource orchestration subjects to the constraints of the system capacity, e.g., limited radio and computing resources, at any time slots.
Hence, we define the second \emph{\textbf{cost}} as $d_k \triangleq \sum\nolimits_{i \in \mathcal{I}}  {a}_{i,k}^{t}, \forall k \in \mathcal{K}$. 
Then, we can satisfy the constraints on the system capacity by ensuring
\begin{equation}
    d_k(\mathbf{a}_t)  \le Z_{k},  \forall k \in \mathcal{K}, t \in \mathcal{T},  \label{rl_const2}
\end{equation}
where $Z_{k}$ is the total amount of the $k$th resource.



On slicing the network, we aim to derive an orchestration policy $\pi^*_\theta$ that minimizes the utilization of cross-domain resources without violating constraints defined by the SLA and system capacity, respectively. Therefore, given a time period $\mathcal{T}$, e.g., 24 hours, we formulate the cross-domain resource orchestration problem in end-to-end network slicing as
\begin{align}
    \mathscr{P}_1: \; &\max \limits_{ \pi_\theta } &\; \mathbb{E}_{\pi_\theta} \left[ \sum\nolimits_{t \in \mathcal{T}} {\gamma^t r(\mathbf{s}_t, \mathbf{a}_t)} \right]  \\ 
      \; & s.t. \; & (\ref{rl_const1}), (\ref{rl_const2}).  \nonumber
\end{align}

The difficulties of solving the above problem is multi-fold.
First, the correlations between the orchestration action $\mathbf{a}_t$ and the slice performance $r(\mathbf{s}_t, \mathbf{a}_t)$ are complicated by the high-dim network state $\mathbf{s}_t$ (e.g., slice traffic, radio channel quality and queuing servers), and thus cannot be accurately represented by mathematical models.
Second, the orchestration action in a time slot would influence not only the current slice performance but also the future network state and performance, which shows a Markov property in end-to-end network slicing.
Therefore, we resort to DRL approaches to deal with the above problem in complex network slicing system.

\vspace{-0.05in}
\section{Constraint-Aware DRL}
\label{sec:design}
In the cross-domain resource orchestration problem, we consider two types of constraints. The first type of constraints are the performance requirements, i.e., the 90-percent end-to-end latency in a day, derived from SLAs (Eq.~\ref{rl_const1}). Since these constraints are on the statistical performance of networks over a period of time, we define them as the statistical constraints. The second type of constraints are the system capacity constraints, e.g., total number of radio resource blocks in the system (Eq.~\ref{rl_const2}). The violation of these constraints depends only on every output resource orchestration action. Hence, we define the second type of constraints as the instantaneous constraints. In this section, we develop a constraint-aware deep reinforcement learning (CaDRL) algorithm that handles the statistical and instantaneous constraints via the policy network and constraint network, respectively.


\subsection{Handling Statistical Constraints}
\label{sec:reward_const}
To handle the statistical constraints in Eq.~\ref{rl_const1}, we utilize Lagrangian primal dual method to incorporate the constraints into the reward with multipliers~\cite{Convex2004Boyd}. The Lagrangian function can be expressed as
\begin{align}
\label{eq:lagrangian_func}
\mathcal{L}(\theta, \lambda)=& \; \mathbb{E}_{\pi_\theta}  \left[ \sum\limits_{t \in \mathcal{T}} \gamma^t \left( {{r}(\mathbf{s}_t, \mathbf{a}_t)} + \sum\limits_{i \in \mathcal{I}} {\frac{\lambda_i}{|\mathcal{T}|}   c_i(\mathbf{s}_t, \mathbf{a}_t)  } \right) \right] + C,
\end{align}
where $C = - \sum\nolimits_{i \in \mathcal{I}} {\lambda_i  {Y}_{i}}$ and $\mathbf{\lambda}$ are multipliers.
The dual function, i.e., the point-wise maximization of Lagrangian with respect to parameters $\theta$ of \emph{policy network}, can be written as 
\begin{align}
\label{eq:dual_func}
    \mathcal{D}\left(\mathbf{\lambda} \right) = \max\limits_{\theta \in (\ref{rl_const2})}  \mathcal{L}(\theta, \mathbf{\lambda}),
\end{align}
where the dual function provides a lower bound on the value of the Lagrangian in Eq.~\ref{eq:lagrangian_func}.
The tighter the bound, the closer the gap between the policy obtained from dual function and the optimal solution of problem $\mathscr{P}_1$.
Thus, the dual problem, which finds the tightest bound, can be expressed as
\begin{align}
\label{eq:dual_prob}
{\min \limits_{ \mathbf{\lambda} \ge 0} \mathcal{D}\left(\mathbf{\lambda} \right)}.
\end{align}
As indicated in \cite{paternain2019constrained}, the duality gap between the primal and dual problem approaches nearly zero if the policy is parameterized by neural networks.
In other words, the problem $\mathscr{P}_1$ can be effectively addressed by alternatively solving the primal problem in Eq.~\ref{eq:dual_func} and the dual problem in Eq.~\ref{eq:dual_prob}.

On one hand, the dual problem is solved by updating the Lagrangian multipliers with sub-gradient descent as
\begin{align}
\label{eq:subgrad}
    \lambda_i^{(m+1)} = \left[\lambda_i^{(m)} - \eta_i  \left( \frac{1}{N |\mathcal{T}|} \sum\limits_{n=0}^{N}  \sum\limits_{t \in \mathcal{T}} \gamma^t c_i(\mathbf{s}_t, \mathbf{a}_t)  - {Y}_{i}\right) \right]^+
\end{align}
where $\eta_i,\forall i \in \mathcal{I}$ are non-negative step sizes and $[x]^+ = \max(0, x)$.
Instead of evaluating $\mathbb{E}_{\pi_\theta}\left[ \sum\nolimits_{t \in \mathcal{T}} \gamma^t  c_i(\mathbf{s}_t, \mathbf{a}_t) \right]$, we approximate it with $N$ times Monte-Carlo sampling.

On the other hand, we note that solving the primal problem, i.e., minimizing the Lagrangian w.r.t. the parameters $\theta$, approximately corresponds to learning a policy with reinforcement learning algorithms such as policy gradient~\cite{sutton2000policy} or actor-critic method~\cite{konda2000actor}.
Thus, we solve the primal problem by updating
\begin{align}
\label{prob:primal}
\theta^{(m+1)} \approx \arg\max \limits_{ \theta \in (\ref{rl_const2})} \mathcal{L}(\theta, \lambda^{(m)}).
\end{align}

\begin{figure}[!t]
	\centering
	\includegraphics[width=3.3in]{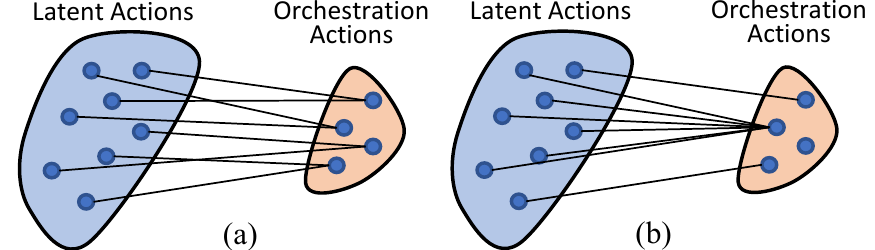}
	\caption{\small Illustration of a) inefficient mapping, b) efficient mapping.}
	\label{fig:constraint_network}
\end{figure}


\subsection{Handling Instantaneous Constraints}
\label{sec:action_const}
When updating \emph{policy network} with parameters $\theta$, we need to maintain that the orchestration actions satisfy the instantaneous constraints in Eq.~\ref{rl_const2} at any time slots.
Since the intrinsic DRL exploration seeks the better reward by exploring the nearby space of actions (e.g., adding noise sampled from Normal distributions), the instantaneous constraints could be easily violated by these deviated actions. 
To address this issue, we design a \emph{constraint network} with parameters $\mu$ to map the latent actions generated by \emph{policy network} to the resource orchestration actions that satisfy the instantaneous constraints. 
The \emph{constraint network} is pre-trained and appended to \emph{policy network} as illustrated in Fig.~\ref{fig:system_design}.


On the designing the \emph{constraint network}, we aim to map the latent actions to final orchestration actions such that the DRL agent can efficiently explore the entire space of the orchestration actions as illustrated in Fig.~\ref{fig:constraint_network}(a). A common mapping method, i.e., clamping~\cite{dalal2018safe}, may result in an inefficient mapping in which a disproportional number of latent actions are mapped to a single orchestration action as shown in Fig.~\ref{fig:constraint_network}(b). With the clamping method, the latent actions that not satisfying the instantaneous constraints are mapped to the boundary of the space defined by the constraints. When the DRL is exploring actions outside the constraint space, the random action exploration mechanism fails as all these explored actions are clamped. As a result, the efficiency of exploring the action space is decreased, and thus degrading the performance of DRL. 

To achieve an efficient mapping, we formulate the problem of designing the \emph{constraint network} as 
\begin{align}
\label{eq:proj_prob}
{\mathscr{P}_2:\;\;\;\;}&{\min \limits_{ \mu} \;\; \; \mathbb{E}_{\mu} \left[D_{KL}\left( p(\mathbf{{a}})|| q(\mathbf{\hat{a}})   \right) \right]}\\
{}&{s.t.\;\;\;\; \mathbb{E}_{\mu} \left[d_k(\mathbf{\hat{a}}) \right] \le Z_{k},  \forall k \in \mathcal{K},}  \nonumber
\end{align}
where $D_{KL}$ is the Kullback-Leibler divergence~\cite{kullback1951information}, $p(\mathbf{{a}})$ and $q(\mathbf{\hat{a}})$ are the distribution of input actions $\mathbf{a}$ and output actions $\mathbf{\hat{a}}$, respectively. We solve problem $\mathscr{P}_2$ by using the Lagrangian primal-dual method.
Specifically, the training loss of the \emph{constraint network} is designed as
\begin{align}
\label{eq:train_loss}
Loss_\mu = D_{KL}\left( p(\mathbf{{a}})|| q(\mathbf{\hat{a}})   \right)   + \sum\limits_{k \in \mathcal{K}} { \beta_k \left(d_k(\mathbf{\hat{a}}) - Z_{k} \right) },
\end{align}
where $\beta_k, \forall k \in \mathcal{K}$ are Lagrangian multipliers that are updated
\begin{align}
\label{eq:proj_subgrad}
    \beta_k^{(m+1)} = \left[\beta_k^{(m)} + \xi_k \left( A - Z_{k} \right) \right]^+, \forall k \in \mathcal{K},
\end{align}
where $A = \max\{ d_k(\mathbf{\hat{a}}_n) , \forall n =1,2,...,N \}$ is the maximum value of the instantaneous constraint under $N$ times Monte-Carlo sampling, and $\xi_k, \forall k \in \mathcal{K}$ are non-negative step sizes.

On training the \emph{constraint network}, the inputs are sampled from the space of the latent actions uniformly.
Here, we aim to build a deterministic mapping from the latent actions to the orchestration actions.
As the pre-trained \emph{constraint network} is appended to \emph{policy network}, we can rewrite the primal problem in Eq.~\ref{prob:primal} as
\vspace{-0.05in}
\begin{equation}
    \theta^{(m+1)} \approx \arg\max \limits_{ \theta} \mathcal{L}(\theta, \lambda^{(m)}).
    \vspace{-0.05in}
\end{equation}

From the perspective of \emph{policy network}, its learning becomes unconstrained, and thus various exploration techniques can be applied without concerning the violation of instantaneous constraints. Since the \emph{constraint network} is based on neural network architecture, which is differentiable, it is compatible with the backpropagation of gradients during \emph{policy network} learning. Hence, the orchestration actions outputted from the \emph{constraint network} are not required to be close to the latent actions outputted from \emph{policy network} in terms of Euclidean distance.

\begin{algorithm}[!t]
	\caption{The CaDRL Algorithm}\label{alg:proposed}

	\KwIn{${Y}_{i}$, $\forall i \in \mathcal{I}$; $Z_{k},\forall k \in \mathcal{K}$; $\eta$, $\xi$. }
	\KwOut{$\pi_{\theta + \mu}$}
	$/**\;constraint\; network\; training\; phase\; **/$\;
	Initialize parameters $\mu$, multipliers $\beta_k, \forall k \in \mathcal{K}$\;
	\For{$m=0,1,...$}
	{
	    Sample batches $\mathcal{B} \gets$ latent actions\;
	    
    	$\mu^{(m+1)} \gets \arg\min \limits_{ \mu} Loss_\mu$\;
    	
    	\If{time to update}
    	{
        	$\beta_k^{(m+1)} \gets Eq.~\ref{eq:proj_subgrad}, \forall k\in\mathcal{K}$\;
        }
    	\If{convergence}
    	{
    	    \textbf{break}\;
    	}
	}
	Append constraint network to policy network\;
	$/**\; offline\; training\; phase\; **/$\;
    Initialize parameters $\theta$, multipliers $\lambda_i, \forall i \in \mathcal{I}$\;
	\For{$m=0,1,...$} 
	{ \label{alg:start}
    	\For{$t=0,1,..., T-1$}
    	{
    	    ${a}_{i,k}^{t}, \forall i\in\mathcal{I}, k \in \mathcal{K}, \gets \pi_{\theta + \mu} $\;
    	    $y_{i}^{t}, \forall i\in\mathcal{I},  \gets $ system\;
    	    $\sum\nolimits_{i \in \mathcal{I}}  {a}_{i,k}^{t}, \forall k\in\mathcal{K}, \gets $ system\;
    	}
    	
    	$\theta^{(m+1)} \gets \arg\min \limits_{ \theta } \mathcal{L}(\theta, \lambda^{(m)})$\; \label{policy_update_in_alg}
    	
    	\If{time to update}
    	{
        	$\lambda_i^{(m+1)} \gets Eq.~\ref{eq:subgrad}, \forall i\in\mathcal{I}$\;
    	}
    	\If{convergence}
    	{   
    	   \textbf{break}\;
    	}
	} \label{alg:end}
	$/**\;online\; learning\; phase\; **/$\;
    Repeat the procedures between Line \ref{alg:start} and Line \ref{alg:end}\;
	\Return{$\pi_{\theta + \mu}$}\;
\end{algorithm}

\subsection{The SafeSlicing Workflow}
\label{subsec:CaDRL}
Deploying Safeslicing mainly consists of three stages: offline constraint network training, offline policy network training, and online policy network learning. 

\textbf{Offline Constraint Network Training.}
The \emph{constraint network} is trained to translate latent actions into resource orchestration actions that satisfy the system constraints.
First, the input batches are sampled uniformly from the latent action space.
Then, these batches are fed into \emph{constraint network} that generates the output actions $\mathbf{\hat{a}}$.
The loss function in Eq.~\ref{eq:train_loss} is calculated, which is used to update \emph{constraint network} accordingly.
After multiple \emph{constraint network} updates, we follow Eq.~\ref{eq:proj_subgrad} to update the multipliers $\beta_k, \forall k \in \mathcal{K}$.
This procedure repeats until the multipliers become stable.

\begin{figure}[t]
	\centering
	\includegraphics[width=3.45in]{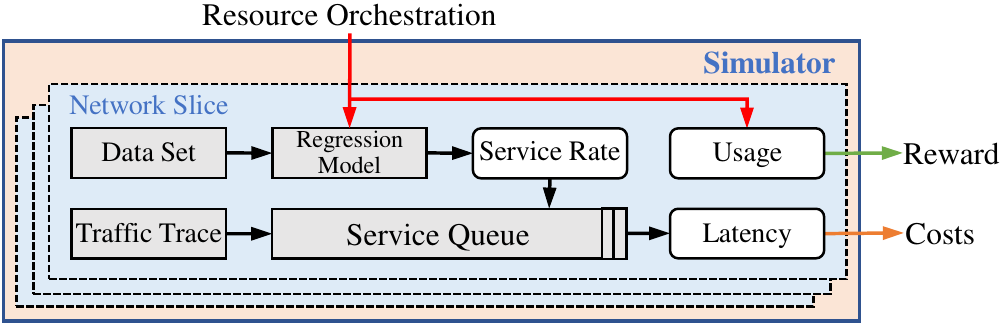}\vspace{-0.05in}
	\caption{\small The design of the simulated environment.}
	\label{fig:simulator}
\end{figure}

\textbf{Offline Policy Network Training.}
We offline train the CaDRL agent with a simulated environment before we conduct the online learning.
The simulated environment is developed to imitate the resource orchestration in the real network by leveraging experimental dataset and exploiting domain knowledge.
As shown in Fig.~\ref{fig:simulator}, we implement a FIFO queue in the environment to simulate the service process for individual slices.
Since the service rates of the queues determine the performance of the slices, it is important to accurately predict the service rater under different resource orchestrations to approximate the performance of the real network. 
Hence, we design a regression model with \emph{scikit-learn}~\cite{scikit-learn} tool in each slice to predict the service rates according to the experimental dataset collected from a real network.
By using the real traffic trace that records the arrival timestamps of user requests, and the predicted service rates, we simulate the enqueuing, serving and dequeuing process of user requests in slices in each time slot.
At the end of a time slot, the application latency of all slices and resource usages are outputted as the costs and rewards, respectively.

During the offline training of \emph{policy network}, we design the agent to observe the following states, i.e., \emph{the average traffic of slices}, \emph{the number of slice users in service queues}, \emph{the number served users of slices}, \emph{the performance of slices} and \emph{the resource usage at the last time slot}. Based on the observed state, the CaDRL agent generates 4-dim resource orchestration for each slice, which are \emph{uplink and downlink physical resource blocks (PRBs) in RAN}, \emph{bandwidth in transport network}, and \emph{the GPU computing resources in edge servers}. The resource orchestration is then fed into the simulated environment shown in Fig.~\ref{fig:simulator}, which updates slice users in service queues according to the traffic trace of slices. The CaDRL agent retrieves the \emph{resource usage} and \emph{end-to-end latency} of slices from the simulated environment as the reward and cost, respectively.


\textbf{Online Policy Network Learning.}
In this phase, we allow the CaDRL agent to directly interact with domain managers, continuously update its \emph{policy network}, and gradually adapt to the real network.
At runtime, the resource orchestrator collects the state space (same with that of offline training) from domain managers and app servers of slices by using TCP sockets.
As the state space is fed into \emph{policy network} and \emph{constraint network} in the CaDRL agent, an orchestration action is generated.
The action is sent to the corresponding domain manager for the action implementation in network infrastructures.
For example, the transport bandwidth allocation of slices are sent to the transport domain manager.
The orchestrator adapts the resource orchestration every 15 minutes. The reward and cost of the orchestration action are defined as the average resource usage and average end-to-end application latency of slices over the 15 minutes period, respectively. The rewards and costs are collected from domain managers and app servers of slices, respectively\footnote{For the deployment in large-scale operational networks, the rewards and costs can be measured via standardized interfaces (e.g., DRM~\cite{nrm}) that support data collections from domain managers and app servers.}.
We define the length of an episode as a day, i.e., 96 transitions per episode\footnote{Although modern cellular networks have relatively long configuration intervals~\cite{marquez2018should}, e.g., 30 mins or 1 hour, we are seeing several efforts on open network architectures that enable real-time or near-RT configurations such as O-RAN~\cite{oran}, which would significantly improve the sampling efficiency for online learning.}.
In the online learning, the CaDRL agent continues exploring the action space and learning from interactions toward the optimal policy in the real network.



\section{System Implementation}
\label{sec:implementation}

\begin{figure}[!t]
	\centering
	\includegraphics[width=3.45in]{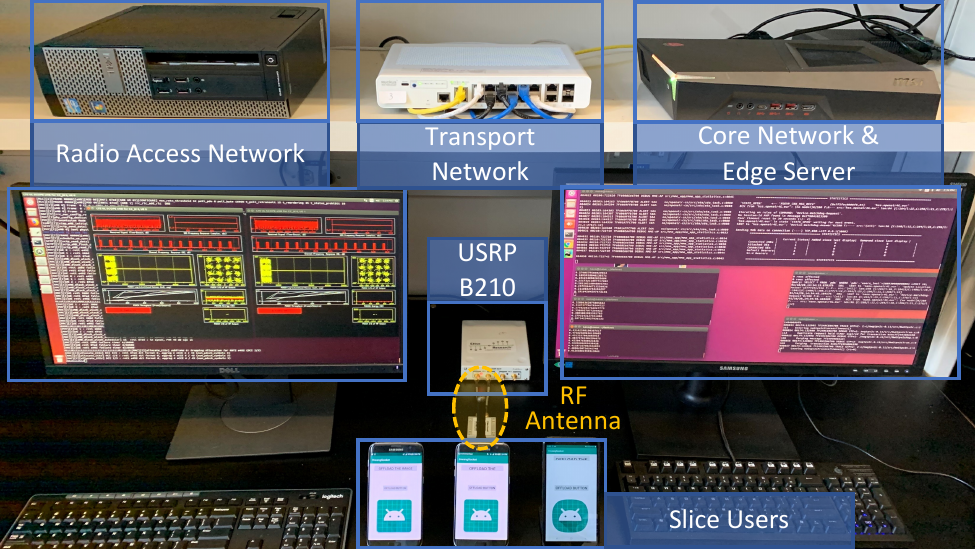}\vspace{-0.05in}
	\caption{\small The overview of system testbed.}
	\label{fig:testbed_overview}
\end{figure}

\textbf{Testbed Setups.}
We prototype \emph{SafeSlicing} on an end-to-end network slicing testbed as depicted in Fig.~\ref{fig:testbed_overview}.
The system testbed is composed of radio access network (RAN) with an eNodeB, transport network with an OpenFlow switch, core network, and edge server with a CUDA GPU.
The details of hardware are summarized in Table~\ref{tbl:hw_detail}.
The domain managers of radio, transport and compute resources are deployed in the computers running eNodeB, a SDN controller and core network, respectively.
The eNodeB is configured with 10MHz (50 PRBs) wireless bandwidth.
The total bandwidth of the transportation network between the eNodeB and the edge server is 100Mbps.
The total amount of the computing resource in the edge server is 51200 CUDA threads.

\textbf{Orchestrator.}
We implement the resource orchestrator by using Python 3.7 and PyTorch 1.40~\cite{paszke2017automatic} and co-locate the resource orchestrator with the core network.
We employ a state-of-the-art reinforcement learning algorithm, i.e., deep deterministic policy gradient (DDPG)~\cite{lillicrap2015continuous}, to update \emph{policy network}.
In particular, we use a 2-layer fully-connected neural network, i.e., [512, 512], in both actor and critic networks in DDPG.
The \emph{constraint network}, which is appended at the end of actor network of DDPG, is implemented with 2-layer fully-connected neural network, i.e., [128, 128].
For both \emph{policy network} and \emph{constraint network}, we use Leaky Rectifier and $sigmoid$~\cite{goodfellow2016deep} activation functions for the intermediate layers and output layer, respectively.
On training the resource orchestrator, we conduct extensive and empirical tuning on the hyper-parameters.
The initial multipliers $\beta_k=0.01, \forall k \in \mathcal{K}, \lambda_i=0.01, \forall i \in \mathcal{I}$. The step sizes of multiplier update $\xi_k=0.1, \forall k \in \mathcal{K}, \eta_i=0.1, \forall i \in \mathcal{I}$.
The learning rates of both actor and critic networks in DDPG and \emph{constraint network} are 5e-4. The batch size is 512.
The discounted factor for cumulative reward is $\gamma=0.99$.
We add a decaying Gaussian noise starts from $\mathcal{N}(0,0.1)$ to $\mathcal{N}(0,0.001)$ on actions during the training phase for balancing the exploitation and exploration.

\begin{table}[!t]
    \small
	\centering
	\vspace{0.1in}
	\begin{tabular}{|c|c|c|}
		\hline
	   \textbf{Component}     &  \textbf{Hardware}  &  \textbf{Software} \\ \hline
   	   UEs     & 3x Smartphones &  Android 7.0\\ \hline
	   eNodeB     & 1x Intel i5 Computer &  OpenAirInterface~\cite{OAI}\\ \hline
	   RF Front-End     & 1x Ettus USRP B210  &  N/A\\ \hline
	   Transport     &   1x OpenFlow Switch &  OpenDayLight~\cite{medved2014opendaylight}\\ \hline
	   Core Network     & 1x Intel i7 Computer &  OpenAir-CN~\cite{openaircn} \\ \hline
	   Edge Server     & 1x Nvidia GTX 1070  &  CUDA 9.0~\cite{nvidia2011nvidia} \\ \hline
	\end{tabular}
	\caption{Details of the system testbed}\label{tbl:hw_detail}
\end{table}

\textbf{Experimental Dataset.}
We collect experimental dataset from the testbed and use it in the simulated environment for training the regression models as shown in Fig.~\ref{fig:simulator}.
We obtain the dataset of the resource orchestration and application latency pairs for each slice, by traversing different combinations of resource orchestration with the grid search method. 
Since the dimension of the resource orchestration space is very high, it is impractical to collect all pairs of the resource orchestration and application latency.
Hence, we collect the data with the granularity of resource allocation equaling $6\%$ for all the resources. 
At the runtime, the resource orchestration generated by the resource orchestrator may not be found in the dataset.
To predict its application latency, we use the adjacent resource orchestration pairs in the dataset to fit the regression model.
Once the model is fitted, the regression model predicts the application latency of the slice under the resource orchestration.

\textbf{Domain Managers.}
\label{sec:domain_manager}
Domain managers are developed to virtualize the physical infrastructures and dynamically implement cross-domain resource orchestrations received from the resource orchestrator into multiple technical domains. In the radio access network, the radio domain manager is designed based on OpenAirInterface~\cite{OAI}, that allocates both uplink and downlink radio resources to slices. The transport domain manager is designed based on an OpenDayLight~\cite{medved2014opendaylight} controller, that allocates bandwidth of the links between radio access network and edge servers through OpenFlow (Southbound API) and RESTful (Northbound API)~\cite{mckeown2008openflow}. The computing domain manager is designed based on CUDA GPU computing platform, which dynamically allocates computing resources, e.g., the number of CUDA threads, to slices.

\section{Performance Evaluation}
\label{sec:evaluation}
In this section, we aim to study 1) how \emph{SafeSlicing} optimizes the network slicing via learning; 2) what's the performance comparison between \emph{SafeSlicing} and existing algorithms; 3) how the design of constraint-awareness impacts the performance of the network slicing; and 4) whether \emph{SafeSlicing} can adapt to the traffic dynamics and learn the actual resource demands of different applications. 

\textbf{Experiment Setups.}
We create three slices on the system testbed, where each slice hosts an application. The application traffic is generated in smartphones that are wirelessly connected to the testbed.
%
For any individual networking and computing resources, the amount allocated to slices should be less than $80\%$, i.e., $Z_{k} =0.8, \forall k \in \mathcal{K}$. We consider the average end-to-end application latency as the main SLA of a slice in the experiments and sets $0.6s$ as the latency requirement, i.e., ${Y}_{i} = 0.6 s, \forall i \in \mathcal{I}$.
Here, the value is selected based on the capability of the testbed and the traffic loads generated from the traffic data trace. A slice is allocated at least $6\%$ of each resource to keep the slice live. 

\textbf{Performance Metrics.}
To evaluate the performance of \emph{SafeSlicing}, we define the following performance metrics. 
The application latency is defined as the average end-to-end latency of the application in an episode. 
The system latency is defined as the average application latency of all slices. 
The resource usage is defined as the average usage of a resource in the system in an episode. 
The system resource usage is defined as the average usages of all resources in the system in an episode.

\textbf{Network Dynamics.}
The network dynamics are composed of varying slice application traffic and different workload of slice applications.
We use \emph{Tsinghua App Usage Dataset} to generate the slice application traffic~\cite{yu2018smartphone}. This dataset consists of traffic traces of nearly 2000 mobile applications. 
We select the social networking category in the dataset and scale down the volume of the requests according to the capability of our end-to-end network slicing testbed. We split the whole selected traffic traces into a train set with 80\% traces and a test set with 20\% traces for training and evaluation, respectively.

To generate different workload of slice applications, we develop three mobile apps that have diverse requirements of networking and computing resources. The first, second and third app have the demands of high-networking \& low-computing, medium-networking \& medium-computing, low-networking \& high-computing, respectively. To simplify the application development, we use a video analytics framework in which video frames are sent to an edge server and analyzed by the YOLO object detection framework~\cite{redmon2016you}. The amount of transmission data, i.e., images with different resolutions, determines the demand of the networking resources, and the complexity of computing model, i.e., YOLO, determines the demand of the computing resources. Hence, the image resolution and computing model of app 1 are 500x500 and tiny YOLOv3, respectively. The image resolution and computing model of app 2 are 300x300 and YOLOv3 416x416, respectively. The image resolution and computing model of app 3 are 100x100 and YOLOv3 608x608, respectively. 
The app server of slices are deployed in the core network, which record the performance of slice users (latency) and report the average slice performance to the CaDRL agent as rewards.

\textbf{Comparison Algorithms.} We compare \emph{SafeSlicing} with the following algorithms:
\underline{Reward Shaping (RS)}: solves the constrained reinforcement learning problem by incorporating the static weighted constraints into the reward function as penalties~\cite{liu2020edgeslice}. After extensive trials with different weights, we select 1.0 as the weight for both statistical and instantaneous constraints.
\underline{Reward Constrained Policy Optimization (RCPO)~\cite{achiam2017constrained}}: is a two-layer training algorithm for constrained reinforcement learning designed based on the Lagrangian primal-dual method. In the upper layer, the Lagrangian multipliers are updated according to the sub-gradient decent method. In the lower layer, the policy network is updated based on the actor-critic method. 


\begin{figure}[!t]
	\centering
	\includegraphics[width=3.48in]{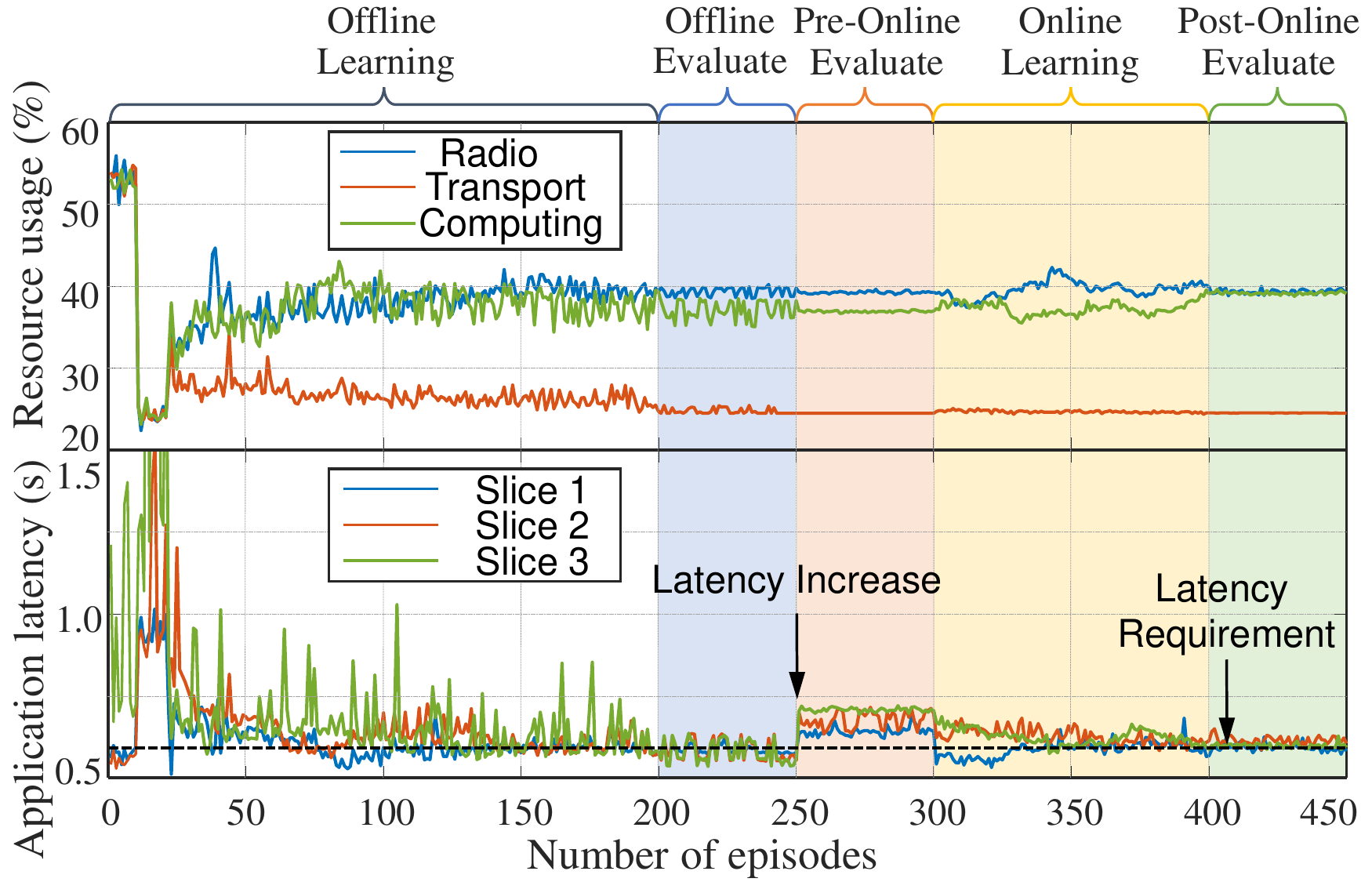}
	\caption{\small Optimizing resource orchestration with \emph{SafeSlicing}.}
	\label{fig:main_training}
\end{figure}


\subsection{Experimental Results}

\subsubsection{Optimizing Network Slicing via learning}
Fig.~\ref{fig:main_training} (a) shows the entire learning-based optimization process of \emph{SafeSlicing}. The offline phase starts from episode 0 to episode 250, in which the first 200 episodes are offline learning, and the episodes 201 to 250 are offline evaluation. After the evaluation, the offline policy is derived. When the offline policy is evaluated in the system (from episodes 251 to 300), the application latency is larger than that in the simulated environment. In other words, the online policy has a poor performance in the real network because of the simulated environment cannot fully reflect the dynamic of the real network. The online learning starts with episode 301 and ends at episode 400. During the online learning, the CaDRL agent interacts with the real network and learns the orchestration policy. In the online evaluation phase, we can see that the application latency of the slices are reduced to around $0.6s$ which is the minimal latency requirement. 


\begin{figure}[!t]
	\centering	
	\includegraphics[width=3.48in,height=1.4in]{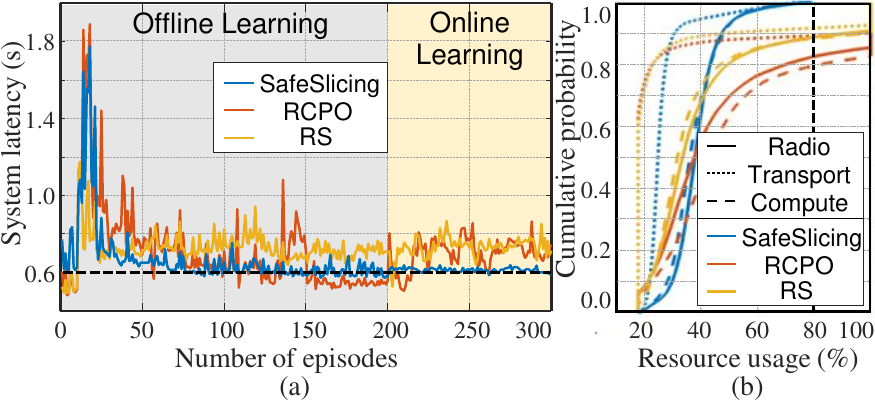}
	\caption{\small The learning performance of different algorithms, a) the system latency; b) the resource usage.}
	\label{fig:alg_latency_usage}
\end{figure}

\subsubsection{Performance Comparison}
We compare the performance of \emph{SafeSlicing} with that of RCPO and RS. Fig.~\ref{fig:alg_latency_usage} (a) shows the system latency of these algorithms. \emph{SafeSlicing} converges to $0.6s$ which the latency requirement defined in the SLA. Meanwhile, both RCPO and RS cannot converge, and the system latency under these algorithms are considerably higher than the latency requirement. This indicates that \emph{SafeSlicing} has better performance in handling the statistical constraints. Fig.~\ref{fig:alg_latency_usage} (b) shows the cumulative probability of the resource usage of these algorithms. We can see that \emph{SafeSlicing} can follow the resource usage constraints, i.e., less than $80\%$ total resources. Both the RCPO and RS violate the resource usage constraints. This proves that \emph{SafeSlicing} can effectively satisfy the instantaneous constraints.

Table~\ref{tbl:results} lists the performance of these algorithms in the post-online evaluation phase. \emph{SafeSlicing} uses the least resources, i.e., $34.3\%\pm 1\%$, and satisfies the system latency constraint, i.e., $0.6s$, with $0\%$ violation of the constraints on the instantaneous resource allocations. Here, \emph{SafeSlicing} maintains the system latency at $0.6s$ with a $\pm 5\%$ error which is tolerant in the SLA. In contrast, although both RCPO and RS consume more resources, they still cannot meet the latency requirement and violate constraints on the resource usages.

\begin{table}[!t]
    \small
	\centering
	\vspace{0.1in}
	\resizebox{0.485\textwidth}{!}{
    	\begin{tabular}{|c|c|c|c|}
    		\hline
    	   \textbf{Metrics}     &  \textbf{SafeSlicing}  &  \textbf{RCPO}  &  \textbf{RS}\\  \hline 
       	   Sys. Resource Usage(\%)    & \textbf{34.3$\pm$0.1} &  42.8$\pm$0.2 & 47.1$\pm$0.9 \\ \hline 
       	   System Latency (s)      & \textbf{0.60$\pm$0.03} &  0.70$\pm$0.25 & 0.71$\pm$0.16\\ \hline
       	   Resource Usage Violation     & \textbf{0.0\%} &  12.7\%  &  12.4\%  \\ \hline
    	   
    	\end{tabular}
	}
	\caption{The performance of different algorithms.}\label{tbl:results}
\end{table}

\begin{figure}[!t]
	\centering
	\includegraphics[width=3.48in, height=1.4in]{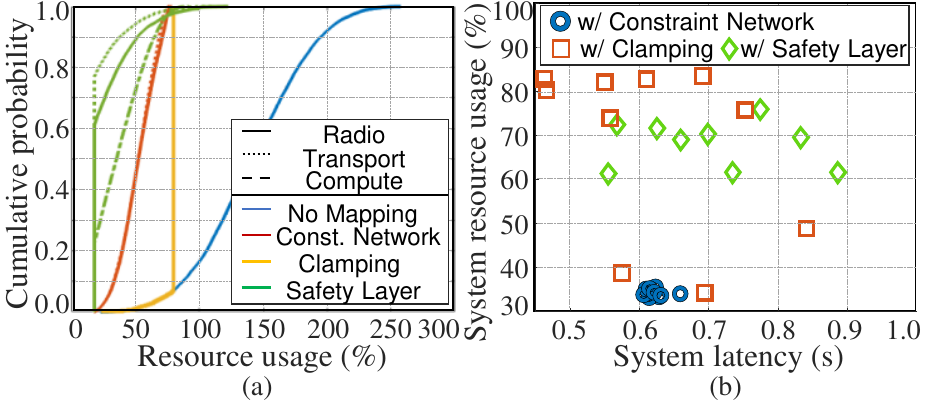} 
	\caption{\small a) The cumulative probability of resource usage under different mappings; b) \emph{SafeSlicing} performance under different mappings.}
	\label{fig:safe_layer}
\end{figure}

\subsubsection{Design of Constraint Awareness}
Although the function of the \emph{constraint network} is simple (i.e., mapping latent actions to orchestration actions), its design determines the performance of the network slicing. Fig.~\ref{fig:safe_layer} (a) shows the cumulative probability of resource usages in the system with different mapping methods. 
Without any mapping, i.e., no constraint-awareness, the resource usages range from $\sim$20\% to $\sim$260\%. With \emph{constraint network}, the range of the resource usage is condensed to $\sim$18\% to $\sim$80\%, which satisfies the constraints on the resource usage, i.e., $Z_{k} =80\%, \forall k \in \mathcal{K}$. Besides, the distribution of the resource usage after the mapping is uniform, which benefits the searching of the orchestration policy. 
The safety layer method introduces an output layer at the end of \emph{policy network} to correct the actions by minimizing the distance between the input and output actions~\cite{dalal2018safe}. The mapping results show that the safety layer method leads to a non-uniform distribution of the resource usage. The clamping method limits the resource usage to the nearest allowable value. Hence, whenever the resource usage exceeds $80\%$, the clamping method maps the usage to $80\%$. 





Fig.~\ref{fig:safe_layer} (b) shows the system latency and resource usage with different mapping methods used in \emph{SafeSlicing}. It shows that the \emph{constraint network} method significantly outperforms the other methods and achieves the shortest system latency with the least resource usage. The reason for the performance gain is that the \emph{constraint network} method allows efficient exploration during \emph{policy network} learning while satisfying the constraints of the system capacity. 


\begin{figure}[!t]
	\centering
	\includegraphics[width=3.0in]{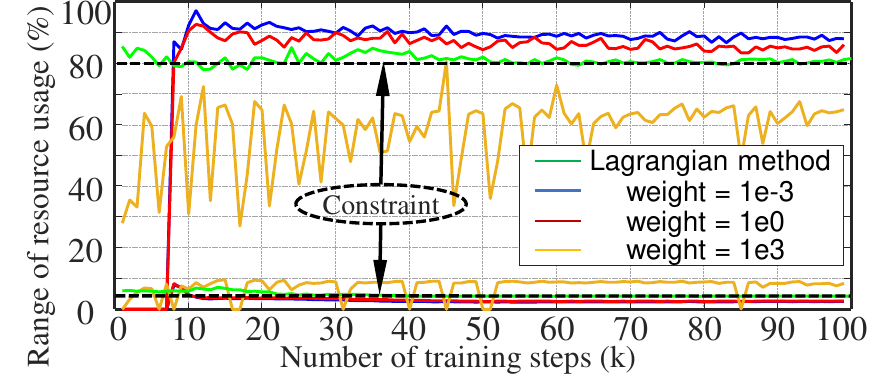} \vspace{-0.05in}
	\caption{\small The impact of weights of constraint network.}
	\label{fig:safe_layer_weights}
\end{figure}

In addition, we show the range of the resource usage during the training of \emph{constraint network} under different static weights in Fig.~\ref{fig:safe_layer_weights}.
It can be observed that these static weighted \emph{constraint networks} either cannot enable the resource usage to approach the upper bound (80\%) and the lower bound (18\%) or violate the resource usage constraints.
In other words, \emph{constraint network} with static weights may generate resource orchestrations that violate the instantaneous constraints and cannot explore some resource orchestrations whose resource usages are close to the boundary.
%
%
In contrast, when the \emph{constraint network} is trained using Lagrangian primal-dual method, the resource usage can approach the upper and lower bounds, which allows the CaDRL agent to explore the entire action space without violating the resource usage constraints.  


\subsubsection{Traffic and Resource Demand Awareness}
\begin{figure}[!t]
	\centering
	\includegraphics[width=3.48in]{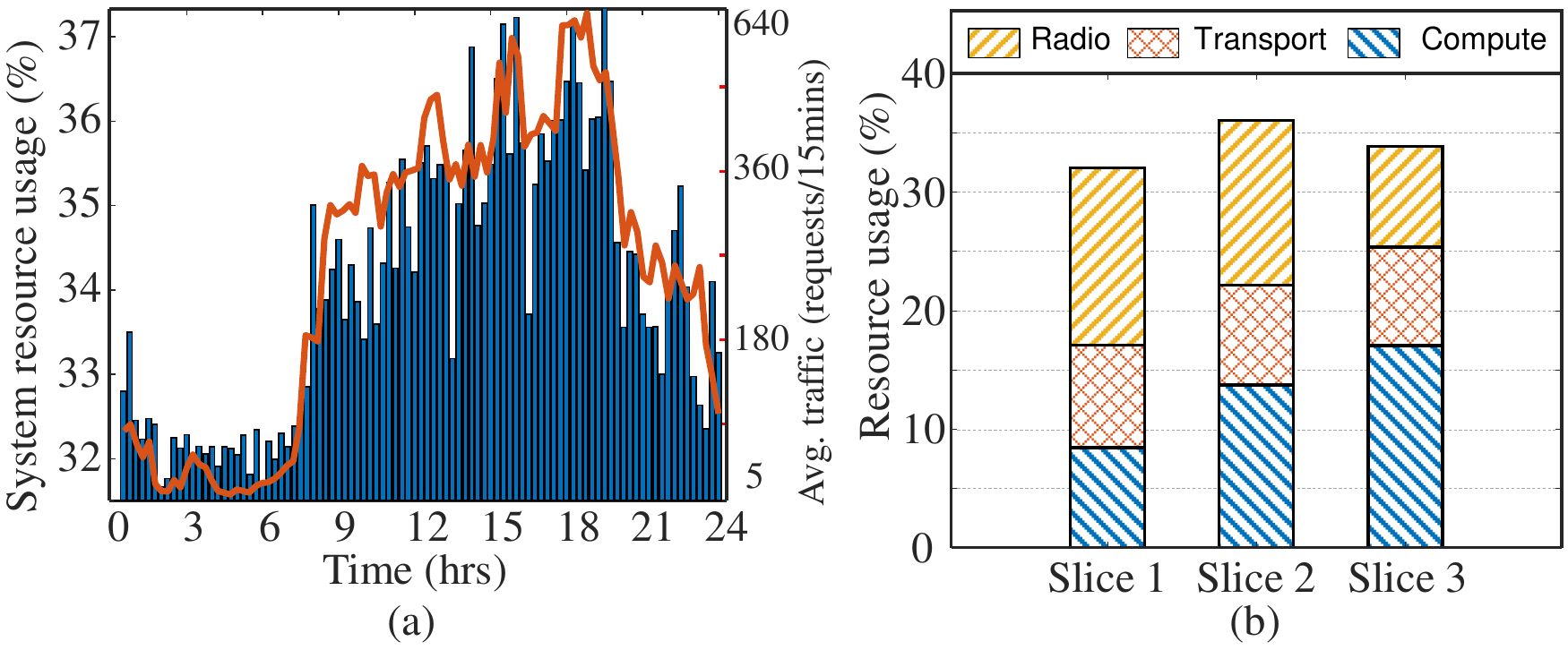}\vspace{-0.05in}
	\caption{\small a) system resource usage vs. average traffic within an episode; b) resource usage of slices.}
	\label{fig:traffic_demand_awareness}
\end{figure}

We show the system resource usage of \emph{SafeSlicing} (post-online) and the corresponded avg. traffic of slices within a episode in Fig.~\ref{fig:traffic_demand_awareness} (a).
The system resource usage track the avg. traffic of slices, which means \emph{SafeSlicing} can react to varying slice traffic and dynamically adjust its resource orchestrations, i.e., traffic awareness, to meet the latency requirement of slices.
Fig.~\ref{fig:traffic_demand_awareness} (b) depicts the resource usage of \emph{SafeSlicing} for slices. It shows that the resource usage in slices align with the resource demands of their application, i.e., resource demand awareness. 
For example, slice 1 and slice 3 adopt the highest frame resolution (500x500) and computation model (YOLOv3 608x608), respectively.
As a result, slice 1 and slice 3 are allocated the largest radio and computing resources, respectively. For these applications, the transportation network is not the bottleneck, and thus the resource allocations in the transportation network are similar. 


\vspace{-0.05in} 
\section{Related Work}
This work relates to resource management problem in network slicing and machine learning techniques for networking. 

\textbf{Resource Management in Network Slicing:}
Salvat \emph{et. al.}~\cite{salvat2018overbooking} proposed a Benders decomposition based algorithm that manages the resource provisioning of slices to maximize the revenue of mobile operators in an end-to-end network slicing system.
Han \emph{et. al.}~\cite{han2019utility} proposed a utility-based admission control mechanism based on multi-queuing systems to improve the resource efficiency for accommodating heterogeneous slices in network slicing.
D'Oro \emph{et. al.}~\cite{d2020sl} proposed a slice admission and resource provisioning algorithm to instantiate slices for heterogeneous services by considering the interdependencies among multiple domain resources in multi-access edge computing (MEC).
%
%
However, these works consider that the performance of slices are mathematically modeled with closed-form expressions. In contrast, \emph{SafeSlicing} is a model-free approach that is capable of handling the dynamic network slicing system with complicated interdependence among cross-domain resources.

\textbf{Machine Learning for Networking:}
Ayala-Romero \emph{et. al.}~\cite{ayala2019vrain} developed a reinforcement learning based approach that dynamically allocates the coupled computing and radio resources to meet heterogeneous QoS targets in virtualized radio access network (vRAN).
Bega \emph{et. al.}~\cite{bega2019deepcog} proposed DeepCog with deep learning techniques to predict network capacity within individual slices and balance the tradeoff between resource over-provisioning and service request violations.
Liu \emph{et. al.}~\cite{liu2020edgeslice} developed an end-to-end network slicing system and proposed a decentralized reinforcement learning approach to orchestrate cross domain resources to meet service level agreement (SLA) of slices.
Liu \emph{et. al.}~\cite{liu2020constrained} proposed a constrained DRL algorithm by using the Interior-point Policy Optimization (IPO) and clamping for dealing statistical and instantaneous constraints, which is trained offline.
Zhang \emph{et. al.}~\cite{zhang2020onrl} designed OnRL, an online DRL solution within real networks, to improve the performance of real-time mobile video telephony, which fails to maintain various constraints in slicing system.
However, these works follow the offline training and online deployment strategy which suffers from the discrepancy between the simulated and real network.
In contrast, with the novel two-stage design of \emph{SafeSlicing}, we achieve constraint-aware deep reinforcement learning that learns the orchestration policy directly with the real network while maintaining various constraints.
\section{Conclusion}
In this paper, we have designed \emph{SafeSlicing} that allows dynamically optimization of network slicing without violating the SLAs of slices.
To accomplish \emph{SafeSlicing}, we have designed a constraint-aware deep reinforcement learning (CaDRL) algorithm that learns the resource orchestration policy while meeting the practical constraints in two steps, i.e., offline training in a simulated environment and online learning with the real network.
The statistical slice performance constraints are incorporated into the reward function with the Lagrangian primal-dual method and the instantaneous resource orchestration constraints are handled by a novel constraint network.
We have developed a system testbed of \emph{SafeSlicing} with OpenAirInterface LTE network, OpenDayLight SDN network, and CUDA GPU computing platform.
We have conducted extensive experiments to evaluate the performance of \emph{SafeSlicing}, and the results have demonstrated that \emph{SafeSlicing} significantly improves on system performance and adheres to the SLAs in the dynamic network slicing optimization. 
\vspace{0.2in}

\bibliographystyle{IEEEtran}
\bibliography{ref/reference}

\end{document}